\documentclass[aps,pra,showpacs,twocolumn]{revtex4}%,superscriptaddress
\usepackage{bm, amsmath, amsfonts}
\usepackage{graphicx}
\usepackage{color}
\begin{document}

\title{On the exact solvability of the anisotropic central spin model: An operator approach}

\author{Ning Wu}
\email{wun1985@gmail.com}
\affiliation{Center for Quantum Technology Research, School of Physics, Beijing Institute of Technology, Beijing 100081, China}

\begin{abstract}
Using an operator approach based on a commutator scheme that has been previously applied to Richardson's reduced BCS model and the inhomogeneous Dicke model, we obtain general exact solvability requirements for an anisotropic central spin model with $XXZ$-type hyperfine coupling between the central spin and the spin bath, without any prior knowledge of integrability of the model. We outline the basic steps of the usage of the operator approach, and pedagogically summarize them into two \emph{Lemmas} and two \emph{Constraints}. Through a step-by-step construction of the eigen-problem, we show that the condition $g'^2_j-g_j^2=c$ naturally arises for the model to be exactly solvable, where $c$ is a constant independent of the bath-spin index $j$, and $\{g_j\}$ and $\{g'_j\}$ are the longitudinal and transverse hyperfine interactions, respectively. The obtained conditions and the resulting Bethe ansatz equations are consistent with that in previous literature.
\end{abstract}

\maketitle

\section{Introduction}
\par The Gaudin model (or the central spin model) proposed by Gaudin in 1976~\cite{Gaudin} describes a central spin $\mathbf{S}_0$ interacting inhomogeneously with a noninteracting spin bath composed of $N$ spins $\{\mathbf{S}_j\}$ via the Heisenberg hyperfine coupling. It is given by the Hamiltonian
\begin{eqnarray}\label{Gaudin}
H_{\rm{Gaudin}}=\sum^N_{j=1}g_j\mathbf{S}_0\cdot \mathbf{S}_j,
\end{eqnarray}
where $g_j$ is the coupling strength between the central spin and the $j$th spin in the bath. While proposed more than 40 years ago, the Gaudin model and its related generalizations nowadays play an important role in solid-state based systems, such as an electron trapped in a quantum dot, which is believed to be a promising platform for realizing quantum computation~\cite{Loss2004,BA2013,PRB2016}.
\par It is well known that the Hamiltonian (\ref{Gaudin}) is integrable and admits a Bethe ansatz solution~\cite{Gaudin,BA2013,JMP2002,Ortiz,Stolze2007,PRB2015}, which has a product form generated by acting a set of parameter-dependent creation operators onto a reference state. Letting the ansatz satisfy the Schr\"odinger equation results in the so-called Bethe ansatz equations that determine the parameters appearing in the ansatz. Given the Bethe ansatz, there are several elegant (but also tricky) ways to derive the Bethe ansatz equations. Among these, Gaudin found out a set of mutually commuting operators among which includes the Gaudin Hamiltonian $H_{\rm{Gaudin}}$. Garajeu and Kiss derived these results using the Lie algebra approach~\cite{JMP2002}. Ortiz \emph{et al.} used a generalized Gaudin algebra and derived the Bethe ansatz equations by diagonalizing the associated Gaudin field operators~\cite{Ortiz}.
\par It is also known that there are close relationships among the Gaudin model, the inhomogeneous Dicke model, and Richardson's reduced BCS model~\cite{Gaudin,Rich,PRB2010,RMP,Inverse,JPA}. For example, Gaudin showed that the Bethe ansatz equations for the inhomogeneous Dicke model can be obtained from those of the Gaudin model in the limit of large central spin size~\cite{Gaudin}. Using a pure operator approach based on the commutator scheme, which was first suggested by Richardson, von Delft and co-workers gave a simplified derivation of the Bethe ansatz equation for the reduced BCS model~\cite{delft1,delft2}. Tsyplyatyev \emph{et al}. then used a similar technique to derive the Bethe ansatz equations for the inhomogeneous Dicke model with the help of an auxiliary pure bosonic model~\cite{PRB2010}. The operator approach was also employed in Ref.~\cite{Inverse} to construct exact eigenstates for a general family of pairing models coupled to a single bosonic mode.
%In spite of these developments along this line, it perhaps seems surprising that such an elementary constructive derivation of the Bethe ansatz equations for the Gaudin model is still absent in the literatures.
\par In this work, we will employ the aforementioned operator approach to derive the exact solvability conditions for the anisotropic central spin model described by the Hamiltonian
\begin{eqnarray}\label{H}
H&=&H_0+H_\perp+H_z,\nonumber\\
H_0&=&h(S^z_0-s_0)+\lambda [(S^z_0)^2-s^2_0],\nonumber\\
H_\perp&=&\frac{1}{2}\sum^N_{j=1}g_j(S^+_0S^-_j+S^-_0S^+_j),\nonumber\\
H_z&=&\sum^N_{j=1}g'_j(S^z_0S^z_j-s_0s_j),
\end{eqnarray}
where $h$ is an external magnetic field acting on the central spin $\mathbf{S}_0$, and $\{g_j\}$ ($\{g'_j\}$) are the in-plane (Ising) part of the anisotropic hyperfine coupling constants~\cite{PRB2016}, which are assumed to be all distinct in order to avoid possible breakdown of the Bethe ansatz method for homogeneous couplings~\cite{homo}. The size of the central spin $\mathbf{S}_0$ and the $j$th bath spin $\mathbf{S}_j$ is assumed to be $s_0$ and $s_j$, respectively, which can be either an integer or a half-integer. We also introduced the longitudinal single-ion anisotropy on the central spin with strength $\lambda$. When $s_0=1/2$, we have $(S^z_0)^2=1/4$, so that the  single-ion becomes a constant.  The $c$-number terms $-hs_0,~-\lambda s^2_0$, and $-\sum_j g'_j s_0s_j$ are included to make $H$ satisfy $H|F\rangle=0$, where
\begin{eqnarray}\label{Fstate}
|F\rangle=|s_0;s_1,\cdots,s_N\rangle
\end{eqnarray}
is the highest-weight state with the first index denoting the central spin. Below we will take $|F\rangle$ as the reference state on which the operator string appearing in the Bethe ansatz acts. We define the total angular momentum of the whole system as $\mathbf{L}=\sum^N_{j=0}\mathbf{S}_j$. We call an operator \emph{friendly} if it takes the reference state $|F\rangle$ as an eigenstate. For example, $H$, $S^z_0$, $S^z_j$, and $L_z=\sum^N_{j=0}S^z_j$ are all friendly operators with eigenvalues $0$, $s_0$, $s_j$, and $l_z=\sum^N_{j=0}s_j$, respectively.
%Note that when $s_0$ or $\{s_j\}$ are greater than $1/2$, it is possible that multiple spin flips occur on the same spin.
\par  As can be easily checked, the total magnetization $L_z$ of the central spin and the spin bath is conserved. %, we therefore introduce the Zeeman term and single-ion term on the central spin only. The corresponding terms on the spin bath can be eliminated as constants for fixed $\mathcal{M}$.
Based on this and without any prior knowledge of solvability of this generalized model, we assume in the beginning that the eigenstates of the model are still expressible in terms of the product-form Bethe ansatz. We then apply the Hamiltonian to the ansatz and follow standard procedures of the operator approach, which are summarized as two lemmas and two constraints, with the expectation of eliminating the unwanted non-eigenstate contributions. The two lemmas consist of commuting certain friendly operators through some operator strings that induce spin flips to hit the reference state $|F\rangle$, which results in a series of complicated commutators and a simpler term proportional to the eigenvalue of the friendly operator. The resulting complicated commutators are then simplified by invoking proper constraints on the parameters appearing in the Bethe ansatz.
\par As we will see, in order to eliminate the unwanted terms consistently, the following condition
\begin{eqnarray}\label{ggc}
g'^2_j-g^2_j=c,~\forall j
\end{eqnarray}
should be imposed, where $c$ is a constant independent of the bath-spin index $j$.
This condition turns out to cover several well-established anisotropic central spin models which act as mutually commuting Gaudin operators in the construction of Bethe ansatz solutions of various Gaudin-like models.
%This condition covers several previously established integrable models including the isotropic Gaudin model~\cite{Gaudin}, the anisotropic Gaudin model with trigonometric, hyperbolic~\cite{PRB2015}, and inverse quadratic couplings~\cite{NPB2017}.
\par It should be noted that the exact solvability condition given by Eq.~(\ref{ggc}) was previously derived by Ortiz \emph{et al.} using a generalized Gaudin algebra~\cite{Ortiz}. However, it seems that Eq.~(\ref{ggc}) is not a well-known result to common readers. Although both Eq.~(\ref{ggc}) and the operator approach used in the present work were generally known in previous literatures, it remains interesting to show how the exact solvability conditions follow from the Bethe ansatz in a relatively straightforward and elementary way with the help of the operator approach. The aim of the present work is to present a general route for the application of the operator approach in such kind of exactly solvable models in a pedagogical way, and to make the method accessible for a wider audience.
\section{The Bethe ansatz wavefunction}
\par Though the total magnetization $L_z$ is a good quantum number, the Hamiltonian (\ref{H}) is generally not exactly solvable unless specific constraints are imposed on the coupling constants. Nevertheless, we will assume in the beginning that the eigenstates of $H$ could still be constructed via a Bethe ansatz of direct product form, due to the conservation of $L_z$. By eliminating the ``unwanted terms" arising from the application of the Hamiltonian, we will see how the condition given by Eq.~(\ref{ggc}) arises naturally by a step-by-step construction of the eigen-problem. Taking $|F\rangle$ as the reference state, we wish to find out an eigenstate $|\Psi_M\rangle$ of $H$ in the $M$-subspace spanned by totally $M$ spin flips. In order to do so, we introduce $M$ (independent) collective spin lowering operators
\begin{eqnarray}\label{Bq}
B^-_q=\sum^N_{l=0}A_{ql}S^-_l,~q=1,2,\cdots,M
\end{eqnarray}
where $\{A_{ql}\}$ are the (not necessarily independent, as we will see) parameters to be determined by letting the following (unnormalized) Bethe ansatz
\begin{eqnarray}\label{product}
|\Psi_M\rangle=P^M_1|F\rangle,%~with~P^M_1\equiv \prod^M_{q=1}B^-_q,
\end{eqnarray}
satisfy the Schr\"odinger equation
\begin{eqnarray}\label{Sch}
H|\Psi_M\rangle=E_M|\Psi_M\rangle.
\end{eqnarray}
Here,
\begin{equation}
P^n_m\equiv
\begin{cases}
 \prod^n_{q=m}B^-_q&m\leq n,\\
 1&m>n,\\
\end{cases}
	\label{Pmn}
\end{equation}
%\begin{eqnarray}\label{Pmn}
%P^n_m\equiv\cases{
% \prod^n_{q=m}B^-_q, & $m\leq n$ \cr
%  1, & $m>n$ \cr
%},
%\end{eqnarray}
%\begin{eqnarray}\label{Pmn}
%P^n_m\equiv
%\begin{cases}
%\prod^n_{q=m}B^-_q,&~m\leq n\\
%1,&~m>n
%\end{cases}
%\end{eqnarray}
and $E_M$ is the corresponding eigenenergy. Since we are not concerned about the normalization of the wavefunction, we shall set $A_{q0}=1,~\forall q$ below.
\par For later use, we also define
\begin{eqnarray}\label{Pmn}
P^{n,(l)}_m&\equiv& P^{l-1}_mS^-_0P^n_{l+1},~(m\leq l\leq n).
%P^{n,(l_1,l_2)}_m&\equiv& P^{l_1-1}_mS^-_0P^{l_2-1}_{l_1+1}S^-_0P^n_{l_2+1},~(m\leq l_1<l_2\leq n).\nonumber\\
\end{eqnarray}
Since $B^-_q$ reduces $l_z$ by one, so $M$ is related to $l_z$ via the relation $M+l_z=\sum^N_{j=0}s_j$. Note that the operator string $P^M_1$ can generate multiple spin flips on the same spin when $s_0$ or $s_j$ is larger than $1/2$.

%In general, when the dimension of the $M$-subspace is larger than the number of parameters in the ansatz, it is possible that a general eigenstate in the $M$-subspace cannot be expressed as a product form as in Eq.~(\ref{product}). However, the validity of the Bethe ansatz ensures that any eigenstate can be written in the form of Eq.~(\ref{product}), and hence reduces (if possible) the direct diagonalization problem of exponential complexity to one of polynomial complexity.
\section{An intuitive derivation of the Bethe ansatz equations}
\par In the following, we will closely follow the operator approach employed in Refs.~\cite{PRB2010,Inverse,delft1,delft2} for dealing with the reduced BCS model and the inhomogeneous Dicke model, which involves commutation relations only. The main steps of this procedure consist of two lemmas and two constraints that arise naturally from the step-by-step construction of the eigen-problem. The Leibniz rule
\begin{eqnarray}\label{OI}
[x,y_1y_2\cdots y_n]&=&[x,y_1]y_2\cdots y_n+y_1[x,y_2]y_3\cdots y_n+\cdots\nonumber\\
&&+y_1 y_2\cdots y_{n-1}[x,y_n],
\end{eqnarray}
for arbitrary operators $x,y_1,\cdots$, and $y_n$ will be frequently used below.
\par We start with the computation of the left-hand side of Eq.~(\ref{Sch}). By noting that $H$ is a friendly operator with eigenvalue $0$, we have:\\
\textbf{\underline{Lemma 1}}:
\begin{eqnarray}\label{Tr1}
H|\Psi_M\rangle=[H,P^M_1]|F\rangle=\sum^M_{q=1}P^{q-1}_1[H,B^-_q]P^M_{q+1}|F\rangle,
\end{eqnarray}
which is a direct consequence of Eq.~(\ref{OI}). We thus need to calculate the following commutator
\begin{eqnarray}\label{Cm1}
[H,B^-_q]&=& S^z_0\sum^N_{j=1}( g_j-A_{qj}g'_j )S^-_j \nonumber\\
&&+ S^-_0 \sum^N_{j=1}(A_{qj} g_j-g'_j )S^z_j\nonumber\\
&&- S^-_0[h+\lambda(2S^z_0-1)],
\end{eqnarray}
which can be directly checked from Eq.~(\ref{H}) and Eq.~(\ref{Bq}). As a byproduct of the above equation, we check the condition under which the total angular momentum $\mathbf{L}$ conserves by setting $A_{qj}=1,~\forall j$, for which the collective spin lowering operator $B^-_q$ reduces to the usual spin lowering operator $L^-$. We then have $[H,L^-]=\sum^N_{j=1}(g_j-g'_j )(S^z_0 S^-_j+S^-_0 S^z_j)- S^-_0[h+\lambda(2S^z_0-1)]$, which means that the total angular momentum $\vec{L}$ is conserved only at the isotropic point $g_j=g'_j$ and in the simultaneous absence of the magnetic field $h$ and the single-ion anisotropy $\lambda$. The conservation of total angular momentum at this specific parameter point was already pointed out in the work of Gaudin~\cite{Gaudin}.
\par As mentioned in the Introduction, the commutator $[H,B^-_q]$ given by Eq.~(\ref{Cm1}) could be simplified by imposing certain constraints on $\{A_{qj}\}$. The main spirit is to gather terms containing spin lowering operators of the bath spins, $\{s_j|j=1,2,\cdots,N\}$, and demand their linear combinations taking the form of the collective spin lowering operator $B^-_q$ given by Eq.~(\ref{Bq}). From inspecting the first term on the right-hand side of Eq.~(\ref{Cm1}), we naturally require $(g_j-A_{qj}g'_j )$ to be proportional to $A_{qj}$ with a $j$-independent coefficient, $-\omega_q$ say, which results in the following\\
\underline{\textbf{Constraint 1}}:
\begin{eqnarray}\label{Ctr1}
A_{qj}=\frac{g_j}{g'_j-\omega_q},~(j=1,2,\cdots,N)
\end{eqnarray}
This constraint is significant since it reduces the number of independent parameter from $MN$ to just $M$. The Bethe ansatz wavefunction $|\Psi_M\rangle$ can thus be written out explicitly as
\begin{eqnarray}\label{wfS}
|\Psi_M\rangle=\prod^M_{q=1}\left(S^-_0+\sum^N_{j=1}\frac{g_j S^-_j}{g'_j-\omega_q}\right)|F\rangle.
\end{eqnarray}
 Actually, the $M$ newly introduced parameters $\{\omega_q\}$ play the role of rapidities which need to be solved for in the usual Bethe ansatz language.
\par We now insert Eq.~(\ref{Ctr1}) into Eq.~(\ref{Cm1}) and obtain
\begin{eqnarray}\label{Cm1_f}
[H,B^-_q]%%&=& -\omega_qS^z_0\sum^N_{j=1}A_{qj}S^-_j+S^-_0 (\omega_q\sum^N_{j=1}A'_{qj}S^z_j-h)-\lambda S^-_0(2S^z_0-1)\nonumber\\
%&\equiv&-\omega_qB^-_qS^z_0 +\omega_qS^-_0S^z_0+S^-_0 (\sum^N_{j=1}\tilde{A}_{qj}S^z_j-h)-\lambda S^-_0(2S^z_0-1)\nonumber\\
&\equiv&-\omega_qB^-_qS^z_0 + S^-_0X_q,
\end{eqnarray}
where we have separated out a term proportional to $B^-_q$, and defined the operator
\begin{eqnarray}
X_q\equiv (\omega_q-2\lambda) S^z_0+   \sum^N_{j=1}  \tilde{A}_{qj} S^z_j-(h-\lambda),
\end{eqnarray}
with
\begin{eqnarray}\label{Ap_qj}
\tilde{A}_{qj}\equiv g'_j\left[\left(\frac{g_j}{g'_j}-\frac{g'_j}{g_j}\right)+ \frac{\omega_q}{g_j}\right]A_{qj}.
\end{eqnarray}
The operator $X_q$ does not induce spin flipping and satisfies
\begin{eqnarray}
X_q|F\rangle=x_q|F\rangle,
\end{eqnarray}
with eigenvalue
\begin{eqnarray}\label{xq}
x_q= (\omega_q-2\lambda) s_0+   \sum^N_{j=1}  \tilde{A}_{qj} s_j-(h-\lambda).
\end{eqnarray}
Thus, $X_q$ is a friendly operator by definition.
\par Substituting Eq.~(\ref{Cm1_f}) into Eq.~(\ref{Tr1}), we have
\begin{eqnarray}\label{LHS_wq}
H|\Psi_M\rangle&=&-\sum^M_{q=1}\omega_qP^{q-1}_1B^-_qS^z_0P^M_{q+1}|F\rangle\nonumber\\
&& +\sum^M_{q=1}P^{q-1}_1S^-_0X_qP^M_{q+1}|F\rangle.
\end{eqnarray}
We now observe that both $S^z_0$ and $X_q$ are friendly operators in the above equation, we thus invoke our\\
\underline{\textbf{Lemma 2}}:
\begin{eqnarray}\label{LHS_wq1}
H|\Psi_M\rangle&=&-\sum^M_{q=1}\omega_qP^{q-1}_1B^-_q[S^z_0,P^M_{q+1}]|F\rangle\nonumber\\
&&+\sum^M_{q=1}P^{q-1}_1S^-_0[X_q,P^M_{q+1}]|F\rangle\nonumber\\
&&-s_0\sum^M_{q=1}\omega_q|\Psi_M\rangle +\sum^M_{q=1}x_q P^{M,(q)}_{1}|F\rangle.
\end{eqnarray}
In turn, we further need the following two commutators
\begin{eqnarray}
[S^z_0,P^M_{q+1}]&=&-\sum^M_{p=q+1} P^{M,(p)}_{q+1},
\label{Cm2}
\end{eqnarray}
\begin{eqnarray}
[X_q,P^M_{q+1}]&=& - \sum^M_{p=q+1}(\omega_q-2\lambda) P^{M,(p)}_{q+1}\nonumber\\
&& - \sum^M_{p=q+1} P^{p-1}_{q+1}  \sum^N_{j=1} \tilde{A}_{qj}A_{pj}S^-_j  P^M_{p+1},
\label{Cm21}
\end{eqnarray}
where we have again used Eq.~(\ref{OI}). The second term in $[X_q,P^M_{q+1}]$ seems complicated. As before, the product
$\tilde{A}_{qj}A_{pj}$ is required to be expressible as a linear combination of the $(A_{kj})'s$, namely, we set\\
\underline{\textbf{Constraint 2}}:
 \begin{eqnarray}\label{ApA1}
\tilde{A}_{qj}A_{pj}= \alpha_{p,q} A_{qj}+\beta_{p,q} A_{pj},~(j=1,2,\cdots,N)
\end{eqnarray}
with $\alpha_{p,q}$ and $\beta_{p,q}$ two parameters not dependent on the index $j$. Note that the $(A_{kj})$'s with $k\neq p,q$ do not contribute to the linear combination since $\tilde{A}_{qj}A_{pj}$ only involves indices $q$ and $p$.
\par If Eq.~(\ref{ApA1}) is fulfilled, then Eq.~(\ref{Cm21}) will become
\begin{eqnarray}\label{XPq}
[X_q,P^M_{q+1}]&=&  \sum^M_{p=q+1} [\alpha_{p,q}+\beta_{p,q}-(\omega_q-2\lambda)]P^{M,(p)}_{q+1}\nonumber\\
&& - \sum^M_{p=q+1}  (\alpha_{p,q}P^{p-1}_{q} P^M_{p+1}+\beta_{p,q}P^M_{q+1}).\nonumber\\
\end{eqnarray}
Now let us see what constraints should be imposed on the coupling constants in order to satisfy Eq.~(\ref{ApA1}). Applying Eq.~(\ref{Ctr1}) and Eq.~(\ref{Ap_qj}) in Eq.~(\ref{ApA1}), we obtain, after some manipulation,
\begin{eqnarray}\label{condition}
&& \alpha_{p,q} \omega_p+\beta_{p,q} \omega_q =(g'^2_j-g^2_j)+g'_j (\alpha_{p,q} +\beta_{p,q}  -\omega_q).\nonumber\\
\end{eqnarray}
Note that the left-hand side of Eq.~(\ref{condition}) is independent of $j$, we therefore impose the condition given by Eq.~(\ref{ggc}) with an additional condition
\begin{eqnarray}\label{AAA}
\alpha_{p,q}+\beta_{p,q}=\omega_q,
\end{eqnarray}
so that the commutator given by Eq.~(\ref{XPq}) becomes
\begin{eqnarray}\label{XPq1}
[X_q,P^M_{q+1}]&=& 2\lambda \sum^M_{p=q+1} P^{M,(p)}_{q+1}\nonumber\\
&&- \sum^M_{p=q+1}  (\alpha_{p,q}P^{p-1}_{q} P^M_{p+1}+\beta_{p,q}P^M_{q+1}),\nonumber\\
\end{eqnarray}
and Eq.~(\ref{condition}) becomes
\begin{eqnarray}\label{condition1}
&&  \alpha_{p,q} \omega_p+\beta_{p,q}  \omega_q =c.
\end{eqnarray}
Solving Eqs.~(\ref{AAA}) and (\ref{condition1}) gives
%\begin{eqnarray}\label{gg0p}
%&&\omega_q(\alpha_{p,q} A_{q0}+\beta_{p,q} A_{p0}-A_{q0}A_{p0})=0,\\
%&&\omega_q(\alpha_{p,q}A_{q0}\omega_p+\beta_{p,q}A_{p0}\omega_q)=cA_{q_0}A_{p0}.
%\end{eqnarray}
%\begin{eqnarray}\label{gg0p}
%&& \frac{\alpha_{p,q}}{A_{p0}}  +\frac{\beta_{p,q}}{A_{q0}}-1 =0,\\
%&&\omega_q(\frac{\alpha_{p,q}}{A_{p0}}\omega_p+\frac{\beta_{p,q}}{A_{q0}}\omega_q)=c.
%\end{eqnarray}
%\begin{eqnarray}\label{gg0p}
%&& \frac{\alpha_{p,q}}{A_{p0}}\omega_p+ \omega_q-\frac{\alpha_{p,q}}{A_{p0}}\omega_q  =\frac{c}{\omega_q}.
%\end{eqnarray}
\begin{eqnarray}\label{gg0p}
\alpha_{p,q} &=&\frac{c-\omega^2_q}{\omega_p-\omega_q},\nonumber\\
\beta_{p,q}&=&-\beta_{q,p}=\frac{\omega_p\omega_q-c}{\omega_p-\omega_q}.
\end{eqnarray}
\par We are now ready to substitute Eq.~(\ref{XPq1}) and Eq.~(\ref{Cm2}) into Eq.~(\ref{LHS_wq1}), and obtain
\begin{eqnarray}\label{LHS_wq2}
H|\Psi_M\rangle
%&=&-\sum^M_{q=1}\omega_qP^{q-1}_1B^-_q[S^z_0,P^M_{q+1}]|F\rangle-s_0\sum^M_{q=1}\omega_q|\Psi_M\rangle\nonumber\\
%&&+\sum^M_{q=1}\omega_qP^{q-1}_1S^-_0[X_q,P^M_{q+1}]|F\rangle\nonumber\\
%&&+\sum^M_{q=1}x_q\omega_qP^{q-1}_1S^-_0P^M_{q+1}|F\rangle\nonumber\\
%&=&-s_0\sum^M_{q=1}\omega_q|\Psi_M\rangle\nonumber\\
%&&+\sum^M_{q=1}\omega_qP^{q-1}_1B^-_q \sum^M_{p=q+1}A_{p0}P^{M,(p)}_{q+1}|F\rangle\nonumber\\
%&&-\sum^M_{q=1}\omega_qP^{q-1}_1S^-_0  \sum^M_{p=q+1}  (\alpha_{p,q}P^{p-1}_{q} P^M_{p+1}+\beta_{p,q}P^M_{q+1})|F\rangle\nonumber\\
%&&+\sum^M_{q=1}x_q\omega_qP^{q-1}_1S^-_0P^M_{q+1}|F\rangle\nonumber\\
%&=&-s_0\sum^M_{q=1}\omega_q|\Psi_M\rangle\nonumber\\
%&&+\sum^M_{p>q}\omega_q   A_{p0}P^{M,(p)}_{1}|F\rangle\nonumber\\
%&&-\sum^M_{p>q}\omega_q(\alpha_{p,q}P_1^{M,(p)}+\beta_{p,q}P_1^{M,(q)})|F\rangle\nonumber\\
%&&+\sum^M_{p=1}x_p\omega_p P^{M,(p)}_{1}|F\rangle\nonumber\\
&=&-s_0\sum^M_{q=1}\omega_q|\Psi_M\rangle+\sum^4_{i=1}|\chi_i\rangle,
\end{eqnarray}
where
 \begin{eqnarray}\label{chi1}
|\chi_1\rangle\equiv \sum_{p>q}\omega_q  P^{M,(p)}_{1}|F\rangle,
\end{eqnarray}
 \begin{eqnarray}\label{chi2}
|\chi_2\rangle\equiv -\sum_{p>q}  (\alpha_{p,q}P^{M,(p)}_{1}+\beta_{p,q}P^{M,(q)}_{1})|F\rangle,
\end{eqnarray}
 \begin{eqnarray}\label{chi3}
|\chi_3\rangle\equiv \sum^M_{p=1}x_p P^{M,(p)}_{1}|F\rangle,
\end{eqnarray}
and
 \begin{eqnarray}\label{chi4}
|\chi_4\rangle\equiv 2\lambda\sum_{p>q}P^{q-1}_1S^-_0P^{M,(p)}_{q+1}|F\rangle.
\end{eqnarray}
\par We see that if we can appropriately choose the parameters such that $|\chi_1\rangle+|\chi_2\rangle+|\chi_3\rangle+|\chi_4\rangle$ vanishes, then we will obtain an eigenstate with eigenenergy
 \begin{eqnarray}\label{EM}
E_M= -s_0\sum^M_{q=1}\omega_q.
\end{eqnarray}
Among these states, $|\chi_4\rangle$ is special since it involves two $(S^-_0)'s$, so we must set $\lambda=0$. Thus, a finite single-ion anisotropy on the central spin will break the exact solvability of the model.
\par In order to eliminate the remaining three terms, we rewrite $|\chi_2\rangle$ as
 \begin{eqnarray}\label{LHS_wq2_line2}
|\chi_2\rangle &=&-\sum_{p>q} \alpha_{p,q}P^{M,(p)}_{1} |F\rangle-\sum_{q>p}   \beta_{q,p}P^{M,(p)}_{1} |F\rangle\nonumber\\
&=&\left(-\sum_{p>q} \alpha_{p,q}P^{M,(p)}_{1} |F\rangle+\sum_{p>q}   \beta_{q,p}P^{M,(p)}_{1} |F\rangle\right)\nonumber\\
&&+\left(-\sum_{q<p}   \beta_{q,p}P^{M,(p)}_{1} |F\rangle-\sum_{q>p}   \beta_{q,p}P^{M,(p)}_{1} |F\rangle\right)\nonumber\\
%&=&-\sum_{p>q}S^-_0  (\omega_q\alpha_{p,q}-\omega_p\beta_{q,p})P^{p-1}_{1} P^M_{p+1} |F\rangle-\sum_{q\neq p}\omega_pS^-_0  \beta_{q,p}P^{p-1}_1P^M_{p+1} |F\rangle\nonumber\\
&\equiv&-|\tilde{\chi}_1\rangle-|\tilde{\chi}_3\rangle,
\end{eqnarray}
where
\begin{eqnarray}\label{chitilde}
|\tilde{\chi}_1\rangle&\equiv& \sum_{p>q}  ( \alpha_{p,q}- \beta_{q,p})P^{M,(p)}_{1}|F\rangle,
\end{eqnarray}
and
\begin{eqnarray}\label{chitilde1}
|\tilde{\chi}_3\rangle&\equiv& \sum^N_{p=1}\sum_{q(\neq p)}   \beta_{q,p}P^{M,(p)}_{1}|F\rangle
\end{eqnarray}
are expected to be identical to  $|\chi_1\rangle$ and $|\chi_3\rangle$, respectively. It is obvious from Eq.~(\ref{AAA}) and Eq.~(\ref{gg0p}) that $|\chi_1\rangle=|\tilde{\chi}_1\rangle$ is satisfied \emph{automatically}. The requirement $|\chi_3\rangle=|\tilde{\chi}_3\rangle$ then leads to
\begin{eqnarray}\label{CD2}
&& x_p-\sum_{q(\neq p)}\beta_{q,p}=0.
\end{eqnarray}
We thus get the expected consistency.
\par By substituting Eq.~(\ref{xq}) and Eq.~(\ref{gg0p}) into Eq.~(\ref{CD2}), we finally arrive at the desired Bethe ansatz equations
%
%
%
% \begin{eqnarray}\label{fn_1}
%s_0A_{p0}+   \sum^N_{j=1} s_j A'_{pj}-\sum_{q(\neq p)}\frac{\omega_q-\frac{c}{\omega_p}}{\omega_q-\omega_p}A_{p0}=0.
%\end{eqnarray}
% \begin{eqnarray}\label{fn_1}
%s_0A_{p0}+   \sum^N_{j=1} s_j g'_j\left[\frac{1}{\omega_p}(\frac{g_j}{g'_j}-\frac{g'_j}{g_j})+ \frac{1}{g_j}\right]A_{pj} -\sum_{q(\neq p)}\frac{\omega_q-\frac{c}{\omega_p}}{\omega_q-\omega_p}A_{p0}=0.
%\end{eqnarray}
% \begin{eqnarray}\label{fn_1}
%s_0+   \sum^N_{j=1} s_j g'_j\left[\frac{1}{\omega_p}(\frac{g_j}{g'_j}-\frac{g'_j}{g_j})+ \frac{1}{g_j}\right]\frac{g_j}{g'_j-\omega_q} -\sum_{q(\neq p)}\frac{\omega_q-\frac{c}{\omega_p}}{\omega_q-\omega_p}=0.
%\end{eqnarray}
% \begin{eqnarray}\label{fn_1}
%s_0+   \sum^N_{j=1}  \left[\frac{1}{\omega_p}(g^2_j-g'^2_j)+ g'_j\right]\frac{ s_j}{g'_j-\omega_q} -\sum_{q(\neq p)}\frac{\omega_q-\frac{c}{\omega_p}}{\omega_q-\omega_p}=0.
%\end{eqnarray}
 \begin{eqnarray}\label{fn_1}
s_0\omega_p+ \sum^N_{j=1} s_j \frac{g'_j\omega_p-c }{g'_j-\omega_p} -\sum_{q(\neq p)}\frac{\omega_p\omega_q-c}{\omega_q-\omega_p}=h,
\end{eqnarray}
where we have used Eq.~(\ref{ggc}). In the absence of the magnetic field and at the isotropic point $g_j=g'_j,~\forall j$, we have $h=c=0$, and the Bethe ansatz equations reduce to the well-known results~\cite{Gaudin}
 \begin{eqnarray}\label{fn_2}
s_0+ \sum^N_{j=1} \frac{g_j s_j }{g_j-\omega_p} -\sum_{q(\neq p)}\frac{\omega_q}{\omega_q-\omega_p}=0.
\end{eqnarray}

\section{Conclusions and Discussions}
\par In this work, we presented an elementary derivation of the Bethe ansatz equations for the anisotropic central spin model. The method we employ is a pure operator approach solely based on commutation relations, and has been successfully applied to Richardson's BCS model and the inhomogeneous Dicke model~\cite{PRB2010,Inverse,delft1,delft2}. By assuming a product-form Bethe ansatz wavefunction, we illustrate the basic ideas and main steps of this elementary approach, which are summarized as two lemmas and two constraints. Within this framework, we show how the exact solvability conditions arise naturally through a step-by-step construction of the eigen-problem.
\par Although the operator approach used here in deriving the known exact solvability conditions Eq.~(\ref{ggc}) is more explicit and elementary than several tricky approaches presented in previous works, it is not obvious whether the approach can be applied to more general models. However, it can at least serves as a primary check of whether a given model can be exactly solvable. For example, if we add an additional term $H'=\sum^N_{j=1}\lambda_j(S^z_j)^2$ that conserves $L_z$ to the original Hamiltonian (\ref{H}), then it can be checked that the term $-\sum^N_{j=1}\lambda_jA_{q j}S^-_j(2S^z_j-1)$ appears in the commutator $[H',B^-_q]$. However, we cannot go further from this point since the term entangles each $S^z_j$ operator with $S^-_j$, so that our ``Constraint 1" cannot be modified to ensure the matching of the required terms.
\par Another concern with the method is whether the exact solutions derived by using the operator approach form a complete basis. We have to say this problem is beyond the scope of the present work. Though the completeness of the Bethe ansatz for the spin-1/2 Rechardson-Gaudin models was established in Ref.~\cite{Links1}, the proof of completeness of the results derived in the present work is a really challenging task.
\\
%, which describes the single-ion anisotropy on the bath spins
\noindent{\bf Acknowledgements:}
We thank J. Links for useful discussions. This work was supported by the NSFC under Grant
No. 11705007 and partially by a startup fund from the Beijing Institute of Technology.
\\

\end{document}